\documentclass[twocolumn,showpacs]{revtex4}
\usepackage{graphicx}
\usepackage{dcolumn}

\begin{document}


\title{Proton-neutron alignment
       in the yrast states of $^{66}$Ge and $^{68}$Ge}

\author{M. Hasegawa$^{1}$, K. Kaneko$^{2}$ and T. Mizusaki$^{3}$}

\affiliation{
$^{1}$Laboratory of Physics, Fukuoka Dental College, Fukuoka 814-0193, Japan \\
$^{2}$Department of Physics, Kyushu Sangyo University, Fukuoka 813-8503, Japan \\
$^{3}$Institute of Natural Sciences, Senshu University, Kawasaki, Kanagawa, 214-8580, Japan
}

\date{\today}

\begin{abstract}

  The $^{66}$Ge and $^{68}$Ge nuclei are studied by means of the shell model
with the extended $P+QQ$ Hamiltonian, which succeeds in reproducing
experimentally observed energy levels, moments of inertia and other properties.
The investigation using the reliable wave-functions predicts $T=0$, $J=9$
one-proton-one-neutron ($1p1n$) alignment in the $g_{9/2}$ orbit,
at high spins ($14_1^+$, $16_1^+$ and $18_1^+$) in these $N \approx Z$
even-even nuclei.  It is shown that a series of the even-$J$
positive-parity yrast states (observed up to $26_1^+$ for $^{68}$Ge)
consists of the ground-state band and successive three bands
with different types of particle alignments (two-neutron, $1p1n$,
two-proton-two-neutron) in the $g_{9/2}$ orbit.

\end{abstract}

\pacs{21.10.-k,21.10.Re,21.60.Cs}

\maketitle

   The study of $N \approx Z$ proton-rich nuclei calls much attention
in the nuclear structure physics, and it is also interesting in a wider
context.  What nuclides exist at the proton drip line?  Are there
special states like isomers which contribute to nucleosynthesis?
Proton-neutron ($pn$) pair correlations are considered to play a key role
in those problems for $N \approx Z$ nuclei.  A lot of effort has been devoted
to the study of the $N \approx Z$ nuclei and the $pn$ pair correlations.
It has explored various aspects of structure such as shape coexistence
and delayed alignment in proton-rich nuclei with $A$=60-80.
The $N \approx Z$ Ge isotopes at the gate to these proton-rich nuclei
have been extensively studied.
The recent development of experimental techniques accomplished detailed
measurements of $^{66}$Ge \cite{Stefanova} and $^{68}$Ge \cite{Ward}.
Our subject is explaining the observed data and clarifying the structure.
Besides this subject, we aim to get a useful effective interaction
for the shell model which is applicable to exploration of the problems
of heavier $N \approx Z$ nuclei.
We have succeeded in reproducing a large number of energy levels
observed in these nuclei.
 Using the wave-functions, we have found a unique phenomenon of particle
alignment which has not been expected in even-even nuclei.
  The particle alignments, which can be considered as breaking away from
the collective $T=1$ or $T=0$ pair correlations caused by rapid rotation,
reveal the features of the $pn$ pair correlations as well as the like-nucleon
pair correlations.
The one-proton-one-neutron ($1p1n$) alignment with $T=0$, $J=2j$ has been
discussed only in odd-odd nuclei. 
In this paper, dealing with $pn$ interactions dynamically in the shell model,
we show unexpected existence of the $T=0$ $1p1n$ alignment
at high-spin yrast states of the $N \approx Z$ even-even nuclei $^{66}$Ge
and $^{68}$Ge.
This is a unique appearance of the $pn$ pair correlations.

   The new experiments for $^{68}$Ge and $^{66}$Ge \cite{Stefanova,Ward}
have found several bands with positive and negative parities up to high spins
($J \le 28$). The new data which display changes in the structure with
increasing spin call our attention to the particle alignments.
The two-nucleon alignment at $J^\pi =8^+$ in $^{68}$Ge and $^{66}$Ge has been
discussed by several authors 
 \cite{Lima,Sound,Petrovici,Boucenna,Chat,Hermkens,Hsieh}.
The calculations based on the deformed mean field approximation
in Ref. \cite{Stefanova} predict simultaneous alignment of protons and neutrons
just after the first band crossing.
   In a previous paper \cite{Kaneko}, we showed that the shell model
with the extended $P+QQ$ interaction in a restricted configuration space
($p_{3/2}$,$f_{5/2}$,$p_{1/2}$,$g_{9/2}$) successfully describes $^{64}$Ge. 
The shell model has advantages that the nuclear deformation is dynamically
determined through nuclear interactions and wave-functions are strictly determined,
which make it possible to calculate physical quantities and to discuss
the structure of bands in detail.
  We carried out large-scale shell model calculations for $^{66}$Ge and
$^{68}$Ge using the calculation code \cite{Mizusaki}.
Results of the calculations explain well all the observed energy levels
and other properties except for the superdeformed band.
We analyze the wave-functions obtained to investigate the structure of
the even-$J$ positive-parity yrast states.

 We first employed the same single-particle energies as those used
for $^{64}$Ge in Ref. \cite{Kaneko}.
The parameters, however, cannot reproduce the relative energies of
the positive and negative parity states in odd-mass Ge isotopes.
We therefore lowered the $g_{9/2}$ orbit toward the $pf$ shell
so that our shell model can reproduce observed level schemes of odd-mass
and even-mass Ge isotopes (and also $^{66}$As) as a whole. 
This was linked with the search for force strengths.
We thus obtained the following set of parameters for the Ge isotopes.
The single-particle energies are
 $\varepsilon_{p3/2} = 0.00$, $\varepsilon_{f5/2} = 0.77$,
 $\varepsilon_{p1/2} = 1.11$ and $\varepsilon_{g9/2} = 2.50$ in MeV.
The strengths of the $J=0$ and $J=2$ pairing, quadrupole-quadrupole and 
octupole-octupole forces are
 $g_0 = 0.262$, $g_2 = 0.0$, $\chi_2 = 0.238$ and $\chi_3 = 0.047$ in MeV.
 The monopole corrections are
 $H^{T=1}_{mc}(p_{3/2},f_{5/2})$=$-$0.3, $H^{T=1}_{mc}(p_{3/2},p_{1/2})$=$-$0.3,
 $H^{T=1}_{mc}(f_{5/2},p_{1/2})$=$-$0.4, $H^{T=1}_{mc}(g_{9/2},g_{9/2})$=$-$0.2
 and $H^{T=0}_{mc}(g_{9/2},g_{9/2})$=$-$0.1 in MeV.

\begin{figure}[t]
\includegraphics[width=8.0cm,height=10.0cm]{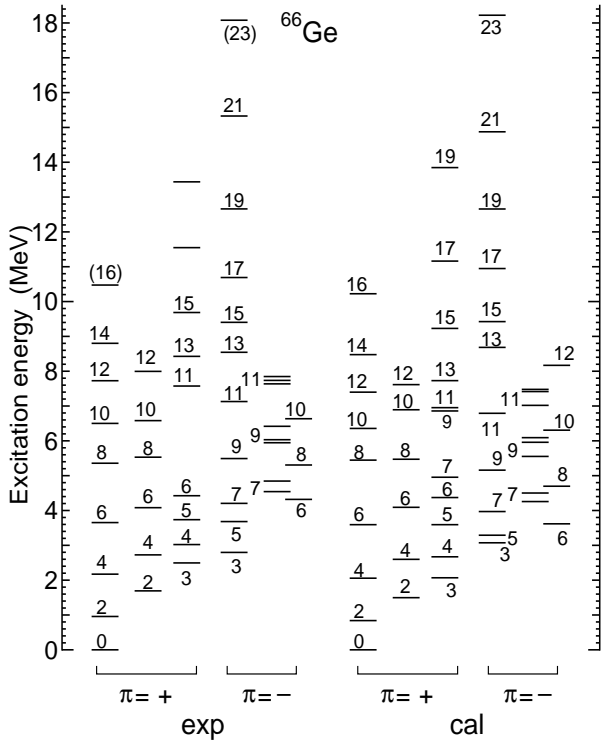}
  \caption{Experimental and calculated energy levels of $^{66}$Ge.}
  \label{fig1}
\end{figure}

\begin{figure}
\includegraphics[width=6.0cm,height=6.2cm]{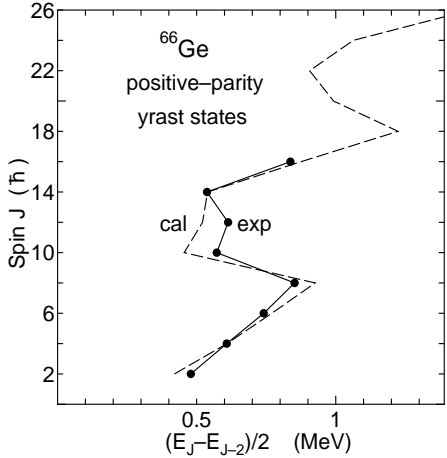}
  \caption{The $J-\omega$ graph for the positive-parity yrast states
           with even $J$ of $^{66}$Ge.}
  \label{fig2}
\end{figure}

   In Fig. \ref{fig1}, we compare energy levels obtained for $^{66}$Ge
with the experimental ones in Ref. \cite{Stefanova}.
The calculations reproduce the several bands of the yrast states
with positive and negative parities observed in $^{66}$Ge.
The agreement between the observed and calculated energy levels is excellent.
The present model reproduces well observed energy levels of $^{68}$Ge
(which are more than twice as many as those of $^{66}$Ge) and
also satisfactorily describes energy levels observed in the odd-mass isotopes
$^{65}$Ge and $^{67}$Ge.
It reproduces the experimental $Q$ moment of the $2_1^+$ state in $^{70}$Ge.
Such a consistent description of both the even and odd Ge isotopes 
has not been reported previously.

  The graph of spin $J$ versus angular frequency $\omega (J)=(E(J)-E(J-2))/2$
(we call it ``$J-\omega$ graph") is useful in seeing the variation
of nuclear structure, because the moment of inertia $J/\omega (J)$ reflects
the competition of various nuclear correlations.
We illustrate the $J-\omega$ graph for the even-$J$ positive-parity
yrast states of $^{66}$Ge, in Fig. \ref{fig2}.   Our model reproduces well
the variation of the experimental moments of inertia.
  The agreement with the experiment is better than that of
the total Routhian surface (TRS) calculations \cite{Stefanova}.
This indicates that our wave-functions are better than those of
 the TRS calculations.
In Fig. \ref{fig2}, the $J-\omega$ graph displays a stable rotation
in the ground-state ($gs$) band up to $8_1^+$ and a sharp backbending
toward $10_1^+$.  The remarkable backbending from $8_1^+$ to $10_1^+$
indicates a structural change there.  The straight line starting
from the $14_1^+$ state is also interesting.

\begin{figure}[b]
\includegraphics[width=7.2cm,height=9.4cm]{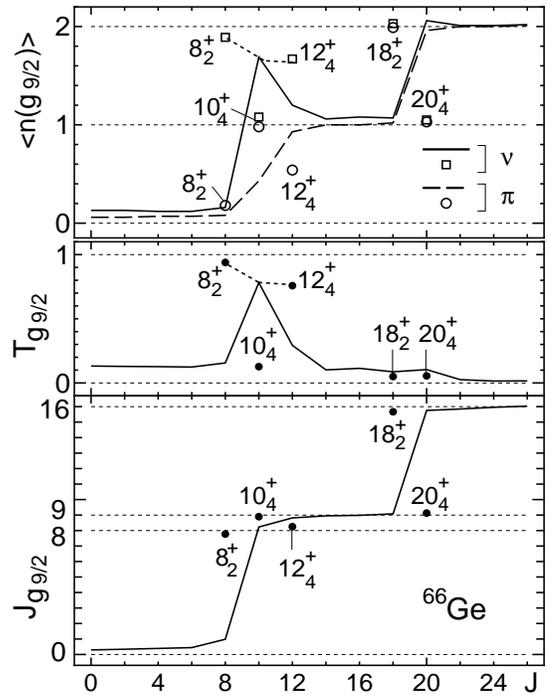}
  \caption{The expectation values $\langle n^\nu _{g9/2} \rangle$
           and $\langle n^\pi _{g9/2} \rangle$ in the upper panel,
           and $T_{g9/2}$ and $J_{g9/2}$ in the lower panel,           
           for the yrast states (lines) and some other states (marks)
           of $^{66}$Ge. }
  \label{fig3}
\end{figure}

  To analyze the wave-functions, we calculated expectation values
of proton and neutron numbers in the four orbits for $^{66}$Ge.
The results show that the nucleon-number expectation values 
$\langle n_a \rangle$ hardly change in the $gs$ band up to $8_1^+$,
which is consistent with the stable rotation
expected from the $J-\omega$ graph.
Above $8_1^+$, the most notable thing is characteristic changes
in the numbers of protons and neutrons occupying the $g_{9/2}$ orbit,
$\langle n^\pi _{g9/2} \rangle$ and $\langle n^\nu _{g9/2} \rangle$.
We illustrate their variations for the yrast states and some other states,
in the upper panel of Fig. \ref{fig3}.  In this figure,
the neutron number in the $g_{9/2}$ orbit ($\langle n^\nu _{g9/2} \rangle$)
increases abruptly  at $10_1^+$.
This change in the wave-function explains the backbending of the experimental
$J-\omega$ graph at $10_1^+$.  However, the abrupt increase of 
$\langle n^\nu _{g9/2} \rangle$ is more remarkable in the $8_2^+$ state,
whereas the proton distribution to the four orbits remains almost
the same as that in the states $0_1^+$ to $8_1^+$.
Since the expectation value $\langle n^\nu _{g9/2} \rangle $ is considered
to be fractional at low energy, the value
$\langle n^\nu _{g9/2} \rangle \approx 2$ (being about integer)
in the $8_2^+$ state suggests the alignment of two neutrons,
$(g_{9/2}^\nu)^2_{J=8,T=1}$.
The large values of $\langle n^\nu _{g9/2} \rangle$  ($\sim 2$)
at $8_2^+$ and $10_1^+$, and a strong $E2$ transition $10_1^+ \rightarrow 8_2^+$
in theory and experiment reveal a similar structure of the two states.
The structural change from $8_1^+$ to $10_1^+$ is probably caused
by the $2n$ alignment coupled to $J=8$, $T=1$ in the $g_{9/2}$ orbit.  
Figure \ref{fig3} suggests a continuation from $8_2^+$ to $10_1^+$.
This situation can be called a ``band crossing", where the $2n$-aligned
band crosses the $gs$ band.
  The present explanation for the $8_1^+$ and $8_2^+$ states is
in agreement with the assignment in the transfer reaction \cite{Boucenna},
the result in the IBM plus a pair treatment \cite{Hsieh} and
the discussion about the kinematic moment of inertia \cite{Stefanova}.

   In Fig. \ref{fig3}, we can see a decrease of the neutron number
$\langle n^\nu _{g9/2} \rangle$  and an increase of the proton number
$\langle n^\pi _{g9/2} \rangle$ from $8_2^+$ to $10_1^+$.
The same trend is clear in the $12_1^+$ state, and then the proton and
neutron numbers become nearly equal to each other in the $14_1^+$ state.
The $14_1^+$, $16_1^+$ and $18_1^+$ states keep nearly integral numbers
$\langle n^\pi _{g9/2} \rangle \approx \langle n^\nu _{g9/2} \rangle \approx 1$.
In our calculation for $^{66}$As using the same Hamiltonian, we had also
almost integral numbers
$\langle n^\pi _{g9/2} \rangle \approx \langle n^\nu _{g9/2} \rangle \approx 1$
for the $J^\pi \ge 9^+$ states of the $T=0$ band.  It is probable
that $^{66}$As has a $T=0$, $J=9$ aligned $1p1n$ pair in these states.
 Similarly, the nearly integral numbers 
$\langle n^\pi _{g9/2} \rangle \approx \langle n^\nu _{g9/2} \rangle \approx 1$
in $^{66}$Ge are presumed to be the signature of the $T=0$, $J=9$ $1p1n$
alignment at the $14_1^+$ states where the $J-\omega$ graph has a notable bend.
   Let us examine it by evaluating the expectation values of
spin and isospin of nucleons in the $g_{9/2}$ orbit.
The lower panel of Fig. \ref{fig3} shows the values
$J_{g9/2}=[\langle ({\hat j}_{g9/2})^2 \rangle +1/4]^{1/2} -1/2$ and
$T_{g9/2}=[\langle ({\hat t}_{g9/2})^2 \rangle +1/4]^{1/2} -1/2$.
 This figure confirms our presumption, telling the following scenario:
The two neutrons in the $g_{9/2}$ orbit outside the $N=Z=32$ central system
align at the $8_2^+$ state and produce the spin $J_{g9/2} \approx 8$
and the isospin $T_{g9/2} \approx 1$.  During the competition
between the $J=8$, $T=1$ $2n$ pair and $J=9$, $T=0$ $1p1n$ pair
in the $10_1^+$ and $12_1^+$ states, the two nucleons in the $g_{9/2}$ orbit
increase the spin and decrease the isospin.
 At last in the $14_1^+$ state where $J_{g9/2} \approx 9$ and
$T_{g9/2} \approx 0$, the $J=9$, $T=0$ $1p1n$ pair overwhelms
 the $J=8$, $T=1$ $2n$ pair.
The superiority of the $J=9$, $T=0$ $1p1n$ pair can be attributed
to the condition that the $T=0$, $J=9$ $pn$ interaction is stronger than
the $T=1$, $J=8$ interaction (note that while the $T=1$, $J=2j-1$ interaction
is repulsive, the $T=0$, $J=2j$ interaction is very attractive
in ordinary effective interactions).
If we set $ \langle (g_{9/2})^2 |V|(g_{9/2})^2:T=0,J=9 \rangle $ zero,
the $1p1n$ aligned states do not become the yrast states, while the $gs$ band
is hardly disturbed.

\begin{figure}[b]
\includegraphics[width=7.2cm,height=7.2cm]{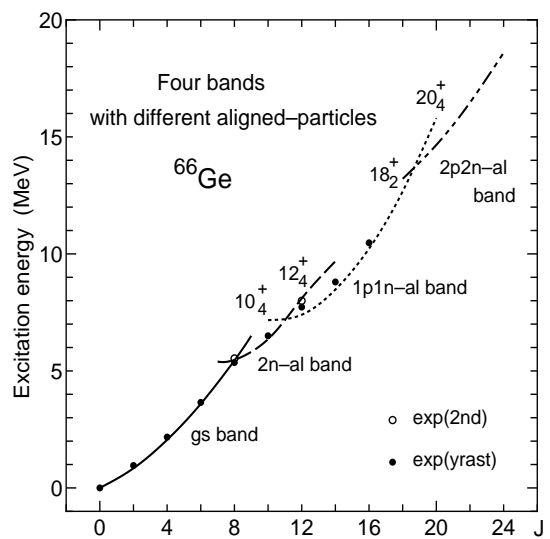}
  \caption{Comparison of the calculated four bands
           with the experimental yrast states for $^{66}$Ge.}
  \label{fig4}
\end{figure}

   To clarify the band crossing near $J=12$, we searched for the $J=10$
member of the $1p1n$ aligned band and the $J=12$ member of the $2n$ aligned
band in our calculations.  Obtained candidates are $10_4^+$ and $12_4^+$,
for which the expectation values $\langle n^\pi _{g9/2} \rangle$, 
$\langle n^\nu _{g9/2} \rangle$, $J_{g9/2}$ and $T_{g9/2}$  are plotted
in Fig. \ref{fig3}.  Figure  \ref{fig3} indicates the band crossing
between $J=10$ and $J=12$.  The TRS calculations \cite{Stefanova}
suggested that the bend at $14_1^+$ of the $J- \omega$ graph is caused
by simultaneous alignment of $2p$ and $2n$.
However, the result presented above disagrees with this suggestion.
As shown in Fig. \ref{fig3}, our model predicts that the simultaneous alignment
of $2p$ and $2n$ takes place at the $18_2^+$ state,
and from $18_2^+$ a band continues to the yrast states
$20_1^+$, $22_1^+$, $24_1^+$ and $26_1^+$.
  Figure \ref{fig3} shows that the $2p2n$ alignment in the $g_{9/2}$ orbit produces
the spin $J_{g9/2} \approx 16$ and the isospin $T_{g9/2} \approx 0$,
which indicates the aligned structure
 $[(g_{9/2}^\pi)^2_{J=8,T=1} (g_{9/2}^\nu)^2_{J=8,T=1}]_{J=16,T=0}$.
  The calculation yields the $20_4^+$ state as the $J=20$ member of
the $1p1n$ aligned band.  The third band crossing takes place
between $J=18$ and $J=20$ in our model.

   What conditions cause such a nearly pure $1p1n$  alignment?
In Ref. \cite{Hase}, we investigated even-mass Ru isotopes around $^{90}$Ru
which is symmetrical to $^{66}$Ge with respect to the particle-hole
transformation in the ($p_{3/2}$,$f_{5/2}$,$p_{1/2}$,$g_{9/2}$) space.
We did not find any sign of the $T=0$ $1p1n$ alignment there,
and could not see a pure $2n$ alignment at the backbending state $8_1^+$
in $^{90}$Ru.
An important thing is that the Fermi level lies at the $g_{9/2}$ orbit itself
in the Ru isotopes but considerably far from the $g_{9/2}$ orbit
in the Ge isotopes.
 The appearance of the nearly pure $2n$ and $1p1n$ alignments in $^{66}$Ge
is based on the condition that the high-spin orbit $g_{9/2}$ is
quite apart from the Fermi level and has the opposite parity to the $pf$ shell.
 Only even-number nucleons are allowed to occupy the $g_{9/2}$ orbit
 after covering the cost of excitation energy from $pf$ to $g_{9/2}$.
 We can expect the $T=0$ $1p1n$ alignment in $N \approx Z$ even-even nuclei
 near the Ge isotopes. 
  It should be also noticed that the residual nucleons in the $pf$ shell
coupled with the aligned $1p1n$ pair with $T=0$, $J=9$ must have
the isospin $T=1$ for the nucleus $^{66}$Ge, while the residual nucleons
coupled with the aligned $2n$ pair with $T=1$, $J=8$ can have the isospins
$T=0$ and $T=1$.  This is confirmed by calculating the isospin of nucleons
in the $pf$ shell.
The different isospin couplings bring about different properties to
the $1p1n$ and $2n$ aligned bands.  The problem is related to the competition
between the $T=1$ and $T=0$ pair correlations in the central system which is
represented by the $pf$ shell in our shell model.
We also calculated the spin of nucleons in the $pf$ shell, ``$J_{pf}$".
The calculated results indicate the approximate alignment of $J_{g9/2}$ and
$J_{pf}$ in the three aligned bands.

  Thus, we have three bands which contain the three types of aligned
nucleons in the $g_{9/2}$ orbit, in addition to the $gs$ band, as shown
in Fig. \ref{fig4}.  
The theoretical bands finely trace the experimentally observed
footprints of the yrast states.  The theory shows a slight deviation
from the experiment near $J=12$, which suggests a stronger coupling
between the $2n$ and $1p1n$ aligned bands.

\begin{figure}
\includegraphics[width=7.4cm,height=7.4cm]{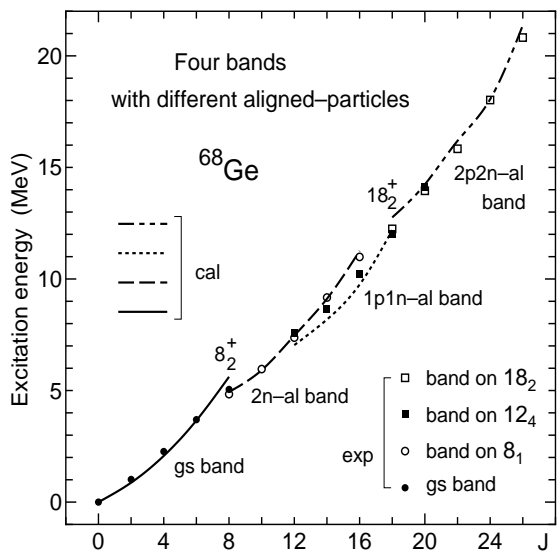}
  \caption{Comparison of the calculated four bands
           with the experimentally observed bands for $^{68}$Ge.}
  \label{fig5}
\end{figure}

   Let us briefly discuss the structure of $^{68}$Ge which has two more
neutrons than $^{66}$Ge.  The present calculations show that the $^{68}$Ge
nucleus has the same structure in the $g_{9/2}$ orbit as that of $^{66}$Ge.
An important difference of $^{68}$Ge from $^{66}$Ge is the backbending
at $8_1^+$ in the $J-\omega$ graph.  We obtained graphs similar to those of
Fig. \ref{fig3}, which explain the band crossing at $J=8$
in terms of the $2n$ alignment in the $g_{9/2}$ orbit.
 The $2n$ aligned band starts from the $8_1^+$ state,
while the $gs$ band continues to the $8_2^+$ state.
This explanation disagrees with the discussions in the particle-rotor model
\cite{Lima} and the VAMPIR calculation \cite{Petrovici}.  We do not have
any sign of the two-proton alignment in our results for $^{68}$Ge as well as
$^{66}$Ge.  There is no significant difference above $J=8$ between 
$^{68}$Ge and $^{66}$Ge.  Also in $^{68}$Ge, we have the same three bands
as those in $^{66}$Ge, as shown in Fig. \ref{fig5}:
the observed band on $8_1^+$ corresponds to the $2n$ aligned band;
the band on $12_4^+$ to the $1p1n$ aligned band; the band on $18_2^+$
to the $2p2n$ aligned band.  The agreement between theory and experiment
is good up to the $26_1^+$ state where the $2p2n$ aligned band terminates. 
 The theory has one deviation from the experiment
with respect to the band crossing at $J=12$.  The $12_1^+$ state
is assigned as the continuation from $10_1^+$ in the experiment \cite{Ward},
while the $12_1^+$ state is the member of the $1p1n$ aligned band
(mixed with the $12^+$ state of the $2n$ aligned band)
in our calculation.  The slightly staggering curves in Fig. \ref{fig5}
show that the coupling between the different bands is stronger
in $^{68}$Ge than in $^{66}$Ge.
We close our discussions by pointing out that the prediction
for the $J>16$ states of $^{66}$Ge in Fig. \ref{fig4} is hopeful.

   In conclusion, we have investigated the mechanism of angular momentum
increase caused by the particle alignments, in the even-$J$ positive-parity
yrast states of $^{66}$Ge and $^{68}$Ge,
using the reliable wave-functions obtained by the successful shell
model calculations.  The investigation has revealed a new feature
that the $T=0$ $1p1n$ alignment in the $g_{9/2}$ orbit takes place
at high spins ($14_1^+$, $16_1^+$ and $18_1^+$)
in the $N \approx Z$ even-even Ge isotopes $^{66}$Ge and $^{68}$Ge.
The three bands with different types of aligned nucleons
in the $g_{9/2}$ orbit successively appear above the $gs$ band
as the spin increases, namely the $2n$ aligned band, the $1p1n$ aligned band
and the $2p2n$ aligned band.



\end{document}